\documentclass[11pt,twoside]{article}

\usepackage{epsfig}

\begin{document}

%\vspace*{4cm} 

%\subsection*{References} 

\title{Route to Room-Temperature Superconductivity\\ from a Practical Point
of View}

\author{A. Mourachkine\footnote{Present address:
Ecole Polytechnique Federal
de Lausanne, 1015 Lausanne, Switzerland }
\\Cavendish Laboratory, University of Cambridge,\\Madingley Road,
Cambridge CB3 0HE, UK}

\date{}

\maketitle

\thispagestyle{myheadings}

\begin{abstract}

To synthesize
a new superconductor which has a critical temperature, $T_c$, exceeding
the room temperature, one needs to know what chemical components to start
with. This chapter presents analysis of experimental data which allow one to
draw a conclusion about components and the structure of a potential
room-temperature  superconductor. The two essential components of a
room-temperature superconductor are large organic molecules (polymers,
tissues) and atoms/molecules which are magnetic in the intercalated state. This
conclusion is fully based on experimental facts known today, and
does not require any
assumptions about the mechanism of room-temperature superconductivity. This,
however, does not mean that to synthesize a room-temperature superconductor
is an easy task.

\vspace{5mm}
\noindent
Never let them persuade you that things are too difficult or impossible.\\
\hspace*{97mm} {\tiny \it ---Sir Douglas Bader}

\end{abstract}

\section{Introduction}

The superconducting state is a state of matter: it is a {\em quantum}
state occurring on a macroscopic scale. As any state of matter,
superconductivity is not a property of isolated atoms, but is a collective effect
determined by the structure of a whole sample. From a classical point of view,
the superconducting state is characterized by two distinctive properties:
{\em perfect electrical conductivity} ($\rho$ = 0) and {\em perfect
diamagnetism} (${\bf B}$ = 0 inside the superconductor, where ${\bf B}$ is
the magnetic field). After the discovery of the phenomenon of superconductivity
in 1911 \cite{1}, humans try to derive a good deal of benefit from its peculiar
properties.
In spite of the fact that superconductivity is a low-temperature phenomenon,
the possibility to use a superconductor at room temperature is an old dream.

It is necessary
first to define the expression ``a room-temperature superconductor''
because some perceive it as a superconductor having a critical temperature
$T_c \sim$ 300 K, others as a superconductor functioning at 300 K. There is
a huge difference between these two cases. From a technical point of view,
superconductors only become useful when they are operated well below
their critical temperature---one-half to two-third of that temperature
provides a rule of thumb. Therefore, for an engineer, a room-temperature
superconductor would be a compound whose resistance disappears
somewhere above 450 K. Such a material could actually be used at room
temperature for large-scale applications. At the same time,
$T_c \sim 350$ K can already be useful for small-scale (low-power)
applications. Consequently, unless specified, the expression ``a
room-temperature superconductor'' will further be used to imply a
superconductor having a critical temperature $T_c \geq$ 350 K.

The invention of the transistor is directly responsible for the way in which
silicon technology has so profoundly changed the world in which we live. The
availability of a room-temperature superconductor may change our lives even
to a greater degree. What technical marvels could we expect to see?

The benefits would range from minor improvements in existing technology to
revolutionary upheavals. All devices made from the room-temperature
superconductor will be reasonably cheap since its use would not involve cooling
cost. Energy savings from
many sources would add up to a reduced dependence on conventional power
plants. Compact superconducting cables would replace unsightly power
lines and revolutionize the electrical power industry. A world with
room-temperature superconductivity would unquestionably be a cleaner
world and a quieter world. Compact superconducting motors would replace
many noisy, polluting engines. Advance transportation systems would lessen
our demands on the automobile. Superconducting magnetic energy storage
would become commonplace. Computers would be based on compact Josephson
junctions. Thanks to the high-frequency, high-sensitivity operation of
superconductive electronics, mobile phones would be so compact that could
be made in the form of an earring. SQUID (Superconducting QUantum
Interference
Device) sensors would become ubiquitous in many areas of technology and
medicine. Room-temperature superconductivity would undoubtedly trigger a
revolution of scientific imagination. The effects of room-temperature
superconductivity would be felt throughout society, including children who
might well grow up playing with superconducting toys.

In the literature, one can find more than 20 papers reporting evidence
of superconductivity near or above room temperature \cite{2}. Most
researchers in superconductivity do not accept the validity of these results
because they cannot be reproduced by others. The main problem
with most of these results is that superconductivity is observed in samples
containing many different phases, and the superconducting
fraction (if such exists at all) of these samples is usually very small. Thus,
superconductivity may exist in these complex materials,
but nobody knows what phase is responsible for its occurrence. In a few cases,
however, the phase is known but superconductivity was observed exclusively
on the surface. For any substance, the surface conditions differ from those
inside the bulk, and the degree of this difference depends on many
parameters, and some of them are extrinsic.

Room-temperature superconductivity was already discussed in a
book \cite{2}. The main purpose of
the book was to show that it is possible to synthesize a room-temperature
superconductor. It was concluded that a room-temperature superconductor
should consist of large organic molecules
(polymers, living tissues) and magnetic atoms/molecules (in the doped state).
This outcome is based on {\bf knowledge} of the mechanism of
high-temperature superconductivity described in the other book \cite{3}. The
mechanism, in its turn, is based on experimental data, mainly, on tunneling
measurements obtained in cuprates. However, the {\bf same} conclusion about
components and the structure of a room-temperature superconductor can
be derived {\em independently}. This is exactly what we are going to do in this
chapter. Thus, to draw a conclusion about components and the structure of a
room-temperature superconductor, it is not necessary to make any
assumptions about the mechanism of room-temperature superconductivity.

The chapter consists of ten sections and is organized as follows. In the
following section, we shall briefly discuss guidelines for materials that
superconduct at high temperatures, presented
by Geballe in 1993. The third section describes the physical
properties of known superconductors. It turns out that all superconductors
can be classified into three groups according to their structural and magnetic
properties. From the analysis of superconducting properties, one can easily
infer that a potential room-temperature superconductor can only belong to one
of these three groups, and not to the other two. Knowledge of the common
physical properties of this group, which are analyzed in the fourth section,
gives an opportunity to know what physical properties should we expect from a
room-temperature superconductor. In the fifth section, we shall discuss the
most important requirements for materials that superconduct at high
temperatures. The principles of superconductivity are briefly discussed in the
sixth section. Bearing in mind the information presented in the first six
sections, an approach to room-temperature superconductivity is proposed
in the seventh section. The presence of bipolarons (bisolitons) in some
non-superconducting polymers and large organic molecules at room
temperature is discussed in the following section. Components and the
structure of a promising room-temperature superconductor are
considered in the ninth section. The chapter ends with conclusions.

\section{Geballe's guidelines}

In 1992, a diverse group of researchers gathered at a two-day workshop in
Bodega Bay (California). They considered the issue of making much higher
temperature superconductors. T. H. Geballe, who attended this workshop,
summarized some guidelines in a two-page paper published in {\em Science}
\cite{4}, that emerged from the discussions:

\begin{itemize}
\item
Materials should be multicomponent structures with more than two sites per
unit cell, where one or more sites not involved in the conduction band can be
used to introduce itinerant charge carriers.

\item
Compositions should be near the metal-insulator Mott transition.

\item
On the insulating side of the Mott transition, the localized states should
have spin-1/2 ground states and antiferromagnetic ordering of the
parent compound.

\item
The conduction band should be formed from antibonding tight-binding states
that have a high degree of cation-anion hybridization near the Fermi level.
There should be no extended metal-metal bonds.

\item
Structural features that are desirable include two-dimensional extended
sheets or clusters with controllable linkage, or both.
\end{itemize}

These hints are mainly based on knowledge of the physical properties of cuprates.
In a sense, one of our tasks in this chapter is to extend these guidelines.

\section{Three groups of superconductors}

The task to synthesize a room-temperature superconductor is a
materials-physics problem. Therefore, it is worthwhile to review superconducting
compounds. Classify all superconducting materials into three groups.
The groups consist of superconductors which are:

1) {\bf three-dimensional and non-magnetic,}

2) {\bf low-dimensional and non-magnetic,} and

3) {\bf low-dimensional and magnetic.}

\noindent
Recently, it was shown that the mechanisms of superconductivity in compounds
of these three groups are different \cite{2,3}. However, this issue is not important
in the context of this chapter. As was mentioned in the Introduction, in order to
draw a conclusion about components and the structure of a room-temperature
superconductor, we shall not try to understand possible mechanisms of
room-temperature superconductivity. Let us briefly review a few superconductors
from these three groups (for more information, see \cite{2}).

\subsection{First group of superconducting materials}

The first group of superconductors incorporates non-magnetic elemental
superconductors and some of their alloys. The superconducting state in
these materials is well described by the BCS theory of superconductivity
\cite{5}. Thus, this group of superconductors includes all
classical, conventional superconductors. The critical temperature of
these superconductors {\bf does not} exceed 10 K. Most of them are type-I
superconductors. As a consequence, superconductors from this group are
not suitable for applications because of their low transitional temperature
and low critical field.

Ironically, many superconductors, discovered mainly before 1986,
were assigned to this group by mistake. In fact, they belong to either the
second or third group of superconductors. For example, the so-called A-15
superconductors, during a long period of time, were considered as
conventional; in reality, they belong to the second group. The so-called
Chevrel phases were first assigned also to the first group; however,
superconductivity in Chevrel phases is of unconventional type, and they
are representatives of the third group of superconductors.

Some elements which superconduct under high pressure belongs either to
the third or the second group of superconductors. Under high pressure,
their crystal structure becomes low-dimensional, often containing
simultaneously two- and one-dimensional substructures. In addition, some of
them are magnetic. For example, under extremely high pressure,
iron exhibits superconductivity which is of unconventional type.
Thus, the number of superconductors in the first group is in fact very
small, and they are not suitable for applications.

\subsection{Second group of superconducting materials}

The second group of superconductors incorporates low-dimensional and
non-magnetic compounds, such as A-15 superconductors, the metal oxide
Ba$_{1-x}$K$_x$BiO$_3$, the magnesium diboride MgB$_2$ and a large number
of other binary compounds. The superconducting state in these materials
is characterized by the presence of two interacting superconducting
subsystems. One of them is low-dimensional and exhibits genuine
superconductivity of unconventional
type, while superconductivity in the second subsystem which is
three-dimensional is often induced by the first one and of the BCS type. So,
superconductivity in this group of materials can be called half-conventional
(or alternatively, half-unconventional). The critical temperature of these
superconductors is {\bf limited} by
$\sim$ 40 K and, in some of them, $T_c$ can be tuned. All of them are
type-II superconductors with an upper critical magnetic field usually
exceeding 10 T. Therefore, many superconductors from this group are
suitable for different types of practical applications.

$\bullet$ Intermetallic compounds of transition metals of niobium (Nb) and
vanadium (V) such as Nb$_3$B and V$_3$B, where B is one of the nontransitional
metals, have the structure of beta-tungsten ($\beta$-W) designated in
crystallography by the symbol A-15. As a consequence, superconductors
having the structure A$_3$B \, (A = Nb, V, Ta, Zr and B = Sn, Ge, Al, Ga, Si)
are called the A-15 superconductors. Nb$_3$Ge has the highest critical
temperature, $T_c =$ 23.2 K. The critical temperature of A-15
superconductors is very sensitive to changes in the 3:1 stoichiometry.
In the crystal structure of the binary A$_3$B compounds,
the atoms B form a body-centered cubic sublattice, while the
atoms A are situated on the faces of the cube forming three sets of
non-interacting orthogonal one-dimensional chains.

$\bullet$
Superconductivity in the metal oxide BaPb$_{1-x}$Bi$_x$O$_3$ was discovered
in 1975, which has a maximum $T_c \simeq$ 13.7 K at $x$ = 0.25. Other
members of this family are BaPb$_{0.75}$Sb$_{0.25}$O$_3$ ($T_c$ = 0.3 K)
and Ba$_{1-x}$K$_x$BiO$_3$ (BKBO). The metal oxide BKBO is an exceptionally
interesting material and the first oxide superconductor without copper with a
critical temperature above that of all the A-15 compounds. Its critical
temperature is $T_c \simeq$ 32 K at $x$ = 0.4. At the moment of writing,
BKBO still exhibits the highest $T_c$ known for an oxide other than the
cuprates. Superconducting BKBO with low potassium content exhibits a
charge-density-wave ordering. The density of charge carriers in BKBO is very
low. Various evidence suggests that the electron-phonon coupling is responsible
for superconductivity in BKBO. A two-band model applied to BKBO accounts very
well for all the available data on BKBO. Acoustic measurements performed in
BKBO show that many physical properties of BKBO are quite similar to those of
the A-15 superconductors.

$\bullet$
In January 2001, magnesium diboride MgB$_2$ was found to superconduct
at $T_c$ = 39 K. At the moment of writing, the intermetallic MgB$_2$ has the
highest critical temperature at ambient pressure among all
superconductors with the exception of superconducting cuprates. The crystal
structure of MgB$_2$ is composed of layers of boron and magnesium,
alternating along the $c$ axis. Each boron layer has a hexagonal lattice similar
to that of graphite. The magnesium atoms are arranged between the boron layers
in the centers of the hexagons. Superconductivity in MgB$_2$ occurs in the
boron layers. The electron-phonon interaction seems to be responsible for the
occurrence of superconductivity in MgB$_2$. The density of states in MgB$_2$
is small. MgB$_2$ has a very low normal-state resistance: at 42 K the resistivity
of MgB$_2$ is more than 20 times smaller than that of Nb$_3$Ge in its normal
state. MgB$_2$ has two energy gaps, $\Delta_L/\Delta_s \simeq$ 2.7.
Seemingly, both the energy gaps have s-wave symmetries: the larger gap is highly
anisotropic, while the smaller one is either isotropic or slightly anisotropic. The
larger energy gap $\Delta_L$ occurs in the $\sigma$-orbital band, while
$\Delta_s$ in the $\pi$-orbital band.

$\bullet$
There are a large number of binary superconductors. Non-magnetic binary
compounds exhibiting high values of $T_c$ and $H_{c2}$ belong to the second
group of superconductors, such as {\em nitrides}, {\em carbides} and
{\em laves phases}. Nitrides and carbides are also known as B1 superconductors.
Metallic AB$_2$ compounds that superconduct are called the laves
phases. Semiconductors, e.g. GeTe and SnTe, also belongs to the second
group of superconductors.

\subsection{Third group of superconducting materials}

The third group of superconductors is the largest and incorporates
superconductors which are low-dimensional and magnetic, or at least, these
compounds have strong magnetic correlations. This is basically the group of
unconventional superconductors.
Superconductors with the highest critical temperature belong to
this group---cuprates show $T_{c,max} \simeq$ 135 K.

In the majority of unconventional superconductors, the magnetic
correlations favor an antiferromagnetic ordering. In contrast to
antiferromagnetic superconductors, ferromagnetic ones usually have a low
critical temperature. The density of charge carriers in these superconductors
is very low. All unconventional superconductors are of type-II. They
have a very large upper critical magnetic field. As a consequence, many
superconductors from this group are used for practical applications.
In this subsection we shall briefly discuss the following compounds from the
third group of superconductors: Chevrel phases, cuprates, charge transfer
organics, fullerides, graphite intercalation compounds, non-organic polymers,
carbon nanotubes, heavy fermions, nickel borocarbides, the strontium
ruthenate, ruthenocuprates, MgCNi$_3$, Cd$_2$Re$_2$O$_7$,
Na$_x$CoO$_2 \cdot y$H$_2$O, hydrides and deuterides.
We start with the so-called Chevrel phases.

\subsubsection{Chevrel phases}

In 1971, Chevrel and co-workers discovered a new class of ternary
molybdenum sulfides, having the general chemical formula
M$_x$Mo$_6$S$_8$, where M stands for a large number of metals and rare
earths (nearly 40), and $x$ = 1 or 2. The Chevrel phases with S substituted by
Se or Te also display superconductivity. Before the discovery of high-$T_c$
superconductivity in cuprates, the A-15 superconductors had the highest values
of $T_c$, but the Chevrel phases were the record holders in exhibiting the
highest values of upper critical magnetic field $H_{c2}$. PbMo$_6$S$_8$ has
the highest critical temperature, $T_c \simeq$15 K, and upper critical magnetic
field, $H_{c2} \simeq$ 60 T.
Superconductivity in the Chevrel phases coexists with antiferromagnetism
of the rare earth elements. For example, a long-range antiferromagnetic
order of the rare earth elements RE = Gd, Tb, Dy and Er in
(RE)Mo$_6$X$_8$, setting in respectively at $T_N$ = 0.84, 0.9, 0.4 and 0.15 K,
coexists with superconductivity occurring at $T_c$ = 1.4, 1.65, 2.1 and 1.85 K,
respectively, where $T_N$ is the N\'eel temperature.
Superconductivity in the Chevrel phases is primarily associated with
the mobile 4$d$-shell electrons of Mo, while the magnetic order involves
the localized 4$f$-shell electrons of the rare earth atoms which occupy
regular positions throughout the lattice.

The crystal structure of Chevrel phases is quite
interesting. These compounds crystallize in a hexagonal-rhombohedral
structure. The building blocks of the Chevrel-phase crystal structure are
the M elements and Mo$_6$X$_8$ molecular clusters. Each
Mo$_6$X$_8$ is a slightly deformed cube with X atoms at the corners, and
Mo atoms at the face centers. The electronic and
superconducting properties of these compounds depend mainly on the
Mo$_6$X$_8$ group, with the M ion having very little effect.

\subsubsection{Cuprates}

A compound is said to belong to the family of copper oxides (cuprates) if it
has the CuO$_2$ planes. Cuprates that superconduct are also called high-$T_c$
superconductors. The first high-$T_c$ superconductor was discovered in
1986 by Bednorz and M\"uller at IBM Zurich Research Laboratory.
The parent compounds of superconducting cuprates are antiferromagnetic
Mott insulators. The cuprates and Na$_x$CoO$_2 \cdot y$H$_2$O (see below)
are the only Mott insulators known to superconduct.
HgBa$_{2}$Ca$_{2}$Cu$_{3}$O$_{10}$
has the highest critical temperature at ambient pressure, $T_c$ = 135 K.
The crystal structure of cuprates is of a perovskite type, and it is highly
anisotropic. Superconductivity in cuprates occur in CuO$_2$ planes.
The CuO$_2$ layers in the cuprates are always separated by layers of other
atoms such as Bi, O, Y, Ba, La etc., which provide the charge carriers into
the CuO$_2$ planes. The ground state of CuO$_2$ planes is antiferromagnetic.
The $T_c(p)$ dependence has a nearly bell-like shape, where $p$ is the hole
(electron) concentration in the copper oxides planes. In spite of the fact that
the structure of cuprates is two-dimensional, the in-plane transport properties
are quasi-one-dimensional. One compound
YBa$_{2}$Cu$_{3}$O$_{6+x}$ has one-dimensional CuO chains.
The majority of superconducting cuprates are hole-doped. The number of
electron-doped cuprates is very limited, e.g. Nd$_{2-x}$Ce$_{x}$CuO$_{4}$
and Pr$_{2-x}$Ce$_{x}$CuO$_{4}$.

\subsubsection{Charge transfer organics}

Organic compounds are usually insulators. It turns out that some of them
superconduct at low temperatures.
The first organic superconductor was discovered in 1979 by Jerome and
Bechgaard: the compound (TMTSF)$_2$PF$_6$ was found to superconduct
below $T_c$ = 0.9 K under a pressure of 12 kbar. TMTSF denotes
tetramethyltetraselenafulvalene, and PF$_6$ is the hexafluorophosphate.
However, the ten years following this discovery saw a remarkable increase
in $T_c$. In 1990, an organic superconductor with $T_c \approx$ 12 K was
synthesized. In only 10 years, $T_c$ increased over a factor 10!
All organic superconductors are layered; therefore, basically they are
two-dimensional. However, the electron transport in some of them is
quasi-one-dimensional.

Figure 1 shows several organic molecules that form superconductors.
In general, they are flat, planar molecules. Among other elements, these
molecules contain sulfur or selenium atoms. In a crystal, these organic
molecules are arranged in stacks. The chains of other atoms or
molecules (PF$_6$, ClO$_4$ etc.) are aligned in these crystals
parallel to the stacks. As an example, the crystal structure of the first
organic superconductor (TMTSF)$_2$PF$_6$, a representative of the
{\em Bechgaard salts}, is schematically shown in Fig. 2. The planar
TMTSF molecules form stacks along which the electrons are most
conducting (the $a$ axis). The chains of PF$_6$ lie between the stacks,
aligned parallel to them. Two molecules TMTSF donate one electron to an
anion PF$_6$.
The separation of charge creates electrons and holes that can become
delocalized to render the compound conducting and, at low temperatures,
superconducting (under pressure).

\begin{figure}[h]
\epsfxsize=0.9\columnwidth
\centerline{\epsffile{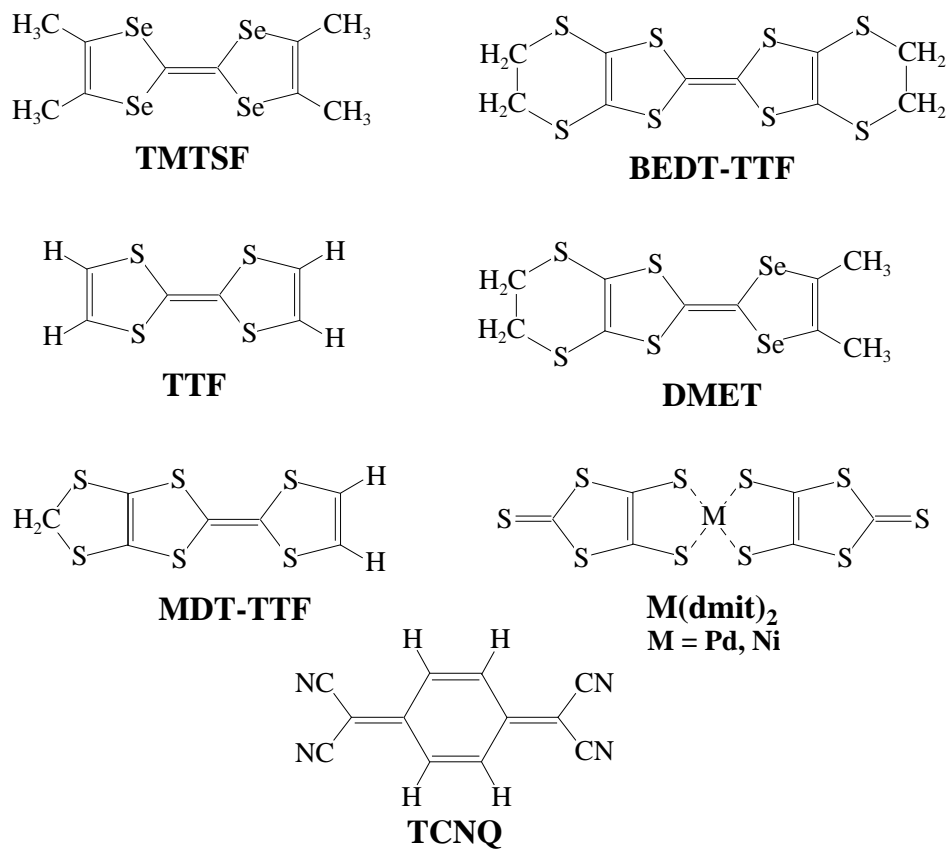}}
\caption{Structure of organic molecules that form superconductors.
Abbreviations of their names are shown below each molecule.}
\end{figure}

\begin{figure}[h]
\epsfxsize=0.87\columnwidth
\centerline{\epsffile{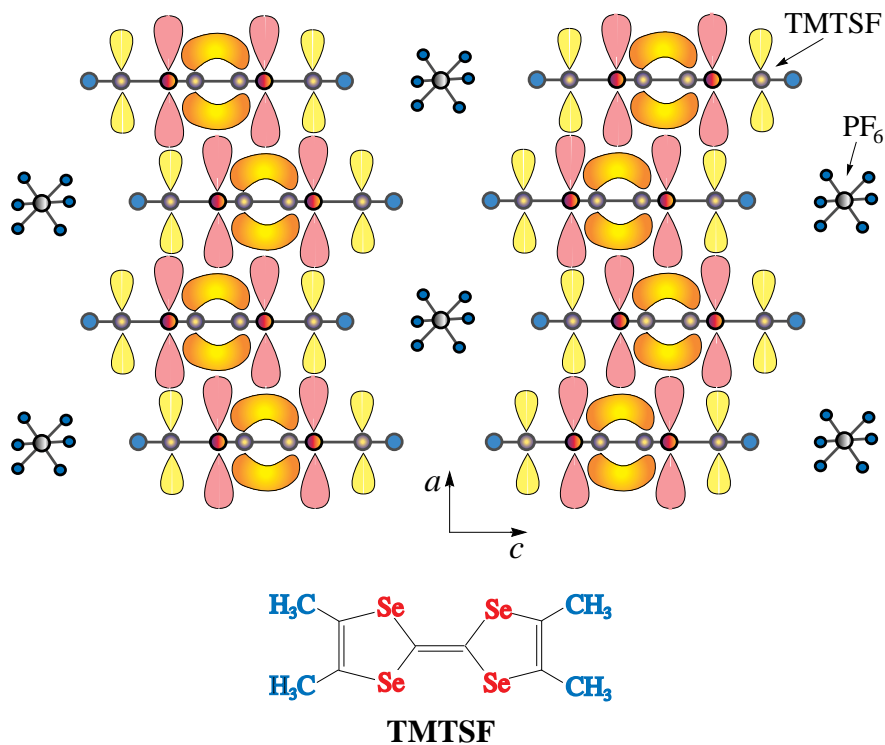}}
\caption{A side view of the crystal structure of the Bechgaard salt
(TMTSF)$_2$PF$_6$. Each TMTSF molecule is shown with the electron
orbitals (the hydrogen atoms are not shown). The chemical structure of
the TMTSF molecule is depicted at the bottom. The organic salt
(TMTSF)$_2$PF$_6$ is the most conductive along the TMTSF stacks
(along the $a$ axis).}
\end{figure}

After 1979, several more organic superconductors of similar structure
were discovered. In all cases, some anion X$^-$ is needed to affect charge
balance in order to obtain metallic properties and, at low temperature,
superconductivity. So, the anions are mainly charge-compensating
spacers; the conductivity is in the organic molecules. There are six
different classes of organic superconductors. Two of them are the
most studied---the Bechgaard salts (TMTSF)$_2$X and the organic salts
(BEDT-TTF)$_2$X based on the compound BEDT-TTF shown in Fig. 1.
BEDT-TTF denotes bis-ethylenedithio-tetrathiafulvalene. The
members of the (BEDT-TTF)$_2$X family exhibit the highest values of
$T_c$, and have a rich variety of crystalline structures. In contrast to
the Bechgaard salts which exhibit quasi-one-dimensional electron
transport, the electronic structure of the BEDT-TTF family is of
two-dimensional nature. The
highest values of $T_c$ are observed in the (BEDT-TTF)$_2$X salts with the
anions X = Cu(NCS)$_2$; Cu[N(CN)$_2]$Br and Cu[N(CN)$_2$]Cl. Their
critical temperatures are respectively $T_c$ = 10.4, 11.6 and 12.8 K. The
first two compounds superconduct at ambient pressure, while the last
one with Cu[N(CN)$_2$]Cl becomes superconducting under a pressure of
0.3 kbar.

Depending on pressure, organic superconductors exhibit a long-range
antiferromagnetic ordering. If, in the phase diagram of the Bechgaard salts,
the superconducting phase evolves out of the antiferromagnetic phase, in
the organic salt $k$-(BEDT-TTF)$_2$Cu[N(CN)$_2]$Br, these two phases
overlap. This suggests that antiferromagnetic fluctuations---short-lived
excitations of the hole-spin arrangements---are important in the
mechanism of unconventional superconductivity in organic salts.

The quasi-two-dimensional organic conductor
$\lambda$-(BETS)$_2$FeCl$_4$, superconductivity is induced by a very
strong magnetic field, 18 $\leq H \leq$ 41 T \cite{2}. The dependence $T_c(H)$
has a bell-like shape with a maximum $T_c \simeq$ 4.2 K near 33 T. At zero
field, this organic compound is an antiferromagnetic insulator below
8.5 K. The other two-dimensional compound,
$\alpha$-(BEDT-TTF)$_2$KHg(NCS)$_4$, at low magnetic fields is a
charge-density-wave insulator. Thus, in these organic salts, the magnetic
and electronic degrees of freedom are coupled. Furthermore, the fact that
the electronic and magnetic properties of organic superconductors strongly
depend on pressure indicates that their electronic, magnetic and crystal
structures are strongly coupled.

\subsubsection{Fullerides}

Historically, any allotrope based on the element carbon has been classed as
organic, but a new carbon allotrope stretches that definition. The pure
element carbon forms not only graphite and diamond but a soccer-ball
shaped molecule containing 60 atoms called  buckminster-fullerene or
buckyball. There are also lower and higher molecular weight
variations such as C$_{20}$, C$_{28}$, C$_{70}$, C$_{72}$, C$_{100}$ and
so forth, which share many of the same properties. The word ``fullerenes''
is now used to denote all these molecules and other closed-cage molecules
consisting of only carbon atoms.

C$_{60}$ was discovered in 1985. In
1991 it was found that intercalation of alkali-metal atoms in solid C$_{60}$
leads to metallic behavior. The alkali-doped fullerenes are called
fullerides. Shortly afterwards, also in 1991, it was
discovered that some of these alkali-doped C$_{60}$ compounds are
superconducting. In fullerides, the maximum critical temperature of 33 K is
observed at ambient pressure in RbCs$_2$C$_{60}$, and $T_c$ = 40 K in
Cs$_3$C$_{60}$ under a pressure of 12 kbar.
Unfortunately, the fullerides are extremely unstable in air, burning
spontaneously, so they must be prepared and kept in an inert atmosphere.
The fullerides are magnetic due to spins of alkali
atoms, which are ordered antiferromagnetically at low temperatures.
The fullerides are electron-doped
superconductors, not hole-doped as cuprates and organic salts.
The values of $H_{c1}$ in the fullerides are very small,
$\sim$ 100--200 Oe, whilst those of $H_{c2}$ are sufficiently large for
electron-doped superconductors, $\sim$ 30--50 T.

\subsubsection{Graphite intercalation compounds}

The first observation of superconductivity in doped graphite goes back to
1965, when superconductivity was observed in the potassium graphite
intercalation compound C$_8$K having a critical temperature of 0.55 K.
Later, superconductivity was observed in other graphite intercalation
compounds (GICs). A single layer of three-dimensional graphite
is defined as a {\em graphene} layer. In GICs, the graphene layers are
separated by the layers of intercalant atoms.
According to the preparation method, the superconducting GICs can be
divided into two subgroups: the stage 1 and stage 2 GICs. The stage 2 GICs
are synthesized in two stages. The structures of the stage 1 and 2 GICs
are different along the
$c$ axis. In the stage 1 GICs, the adjacent intercalant layers are separated
from one another by {\em one} graphene layer, while in the stage 2 GICs, the
neighboring intercalant layers are separated by {\em two} graphene layers.
The stage 1 GICs consist of the binary C$_8$M, ternary C$_4$MHg and
C$_4$MTl$_{1.5}$ compounds, and the stage 2 GICs are represented by the
ternary C$_8$MHg and C$_8$MTl$_{1.5}$, where M = K, Rb and Cs.  In the
superconducting GICs,
as well as in the fullerides, the charge carriers are electrons, not holes.

For binary C$_8$M compounds, the
highest critical temperatures reported for M = K, Rb and Cs are 0.55, 0.15
and 0.135 K, respectively. In the alkali metal amalgam GICs C$_8$KHg and
C$_8$RbHg, the critical temperatures are 1.93 and 1.44 K, respectively.
In the potassium thallium GICs C$_4$KTl$_{1.5}$ and C$_8$KTl$_{1.5}$,
respectively $T_c$ = 2.7 and 1.3 K. With the potassium thallium GICs
excluded, the critical temperature of the stage 2 GICs is in general higher
than that of the stage 1 GICs. Under pressure, the sodium graphite
intercalation compound C$_2$Na superconducts below $T_c \sim$ 5 K.
All the GICs are two-dimensional. In C$_4$KTl$_{1.5}$ which has the
highest $T_c$ at ambient pressure (= 2.7 K), $H_{c2, \bot} \simeq$ 3 T.

At the moment of writing, superconductivity in the isostructural graphite
intercalation compounds C$_6$Yb and C$_6$Ca, with transition temperatures
of 6.5 K and 11.5 K, respectively, was discovered \cite{6}.

\subsubsection{(SN)$_x$ polymer}

(SN)$_x$ is a chain-like inorganic polymer in which sulphur and nitrogen atoms
alternate along the chain. When doped with bromine, it becomes superconducting
below $T_c$ = 0.3 K. Its unit cell contains two parallel
spirals of (SN)$_x$ twisted in the opposite directions. The Br$_3^-$ and
Br$_5^-$ clusters are situated between the (SN)$_x$ spirals.
Superconductivity in (SN)$_x$ was discovered in 1975. It is the first
superconductor found among quasi-one-dimensional conductors and,
moreover, the first that does not contain metallic elements.
The single crystals have a {\em dc} electrical conductivity
of about 1.7 $\times$ 10$^5$ $\Omega^{-1}$\,m$^{-1}$ along the chains, and
the anisotropy is of the order of 10$^3$. A remarkable property of (SN)$_x$
is that it does not undergo a metal-insulator (Peierls) transition at low
temperatures but turns instead into a superconductor below 0.3 K.

\subsubsection{Carbon nanotubes}

In addition to ball-like fullerenes, it is possible to synthesize tubular fullerenes.
By rolling a graphene sheet into a cylinder and capping each end of the cylinder
with a half of a fullerene molecule, a fullerene-derived tubule, one atomic layer,
is formed. Depending on the wrapping angle, one can have three types of the
nanotubes: zigzag, armchair and chiral. The armchair nanotubes are usually
metallic, while the zigzag ones are semiconducting. The carbon nanotubes and
fullerenes have a number of common features and also many differences.
Carbon nanotubes can be viewed as giant conjugated molecules with a conjugated
length corresponding to the whole length of the tube.
The nanotubes have an impressive list of attributes. They can behave like
metals or semiconductors, can conduct electricity better than copper, can
transmit heat better than diamond. They rank among the strongest
materials known, and they can superconduct at low temperatures.

Carbon nanotubes were found by Iijima in 1991 in Japan. In fact,
they were multi-walled carbon nanotubes consisting of several concentric
single-walled nanotubes nested inside each other, like a Russian doll.
Two years later, single-walled nanotubes were seen for the first time.
In 1999, proximity-induce superconductivity below 1 K was
observed in single-walled carbon nanotubes, followed by the observation of
genuine superconductivity with $T_c$ = 0.55 K. In the latter case, the
diameter of single-walled nanotubes was of the order of 14 \AA. Soon
afterwards, superconductivity below $T_c \simeq$ 15 K was seen in
single-walled carbon nanotubes with a diameter of 4.2 \AA.

\subsubsection{Heavy-fermion systems}

This family of superconductors includes superconducting compounds which
consist of one magnetic ion with 4$f$ or 5$f$ electrons (usually Ce or U)
and other constituent or constituents being $s$, $p$, or $d$ electron metals.
The principal feature of these materials is reflected in their name: below a
certain coherence temperature ($\sim$ 20--100 K), the effective mass of
charge carriers in these compounds become gigantic, up to several hundred
times greater than that of a free electron. A large number of heavy fermions
superconduct exclusively under pressure. The $T_c$ values of
superconducting heavy fermions are in general very low; however, the family
of these intermetallic compounds is one of the best examples of highly
correlated condensed matter systems. The crystal structure of
these compounds does not have a common pattern, but varies from case to
case. For example, the crystal structure of the first discovered
superconducting heavy fermions---CeCu$_2$Si$_2$, UBe$_{13}$
and UPt$_3$---is tetragonal, cubic and hexagonal, respectively.

The first heavy fermion, CeCu$_2$Si$_2$, was discovered in 1979 by
Steglich and co-workers, and some time passed before the heavy-fermion
phenomenon was confirmed by the discovery of UBe$_{13}$ and then
UPt$_3$, with critical temperatures of $T_c =$ 0.65, 0.9 and 0.5 K,
respectively. Since then many new heavy-fermion systems that
superconduct at low temperatures have been found.

Probably, the most interesting characteristic of superconducting heavy
fermi-on materials is the interplay between superconductivity and
magnetism. The magnetic ions are responsible for the magnetic properties
of heavy fermions. For example, in the heavy fermions UPt$_3$,
URu$_2$Si$_2$, UCu$_5$ and CeRhIn$_5$, magnetic correlations lead to an
itinerant spin-density-wave order, while, in UPd$_2$Al$_3$ and
CeCu$_2$Si$_2$, to a localized antiferromagnetic order. In the latter two
heavy fermions, the antiferromagnetic order appears first, followed by the
onset of superconductivity. In these compounds, as well as in other
superconducting heavy fermions with long-range antiferromagnetic order,
the N\'eel temperature is about $T_N \sim 10 T_c$. For instance,
in CeRh$_{0.5}$Ir$_{0.5}$In$_5$ and CeRhIn$_5$, the bulk superconductivity
coexists {\em microscopically} with small-moment magnetism
($\leq 0.1\mu _{B}$). In the heavy fermion CeIrIn$_5$, the onset of a
small magnetic field ($\sim 0.4$ Gauss) sets in exactly at $T_c$.

Recently, superconductivity was discovered in PuCoGa$_5$, the first
superconducting heavy fermion based on plutonium. What is even more
interesting is that the superconductivity survives up to an astonishingly
high temperature of 18 K.

It was a surprise when in 2000 the coexistence of
superconductivity and ferromagnetism was discovered in an alloy of
uranium and germanium, UGe$_2$. At ambient pressure, UGe$_2$ is known
as a metallic ferromagnet with a Curie temperature of $T_C$ = 53 K.
However, as increasing pressure is applied to the ferromagnet, $T_C$ falls
monotonically, and appears to vanish at a critical pressure of $P_c \simeq$
16--17 kbars. In a narrow range of pressure below $P_c$ and thus
{\em within} the ferromagnetic state, the superconducting phase appears
in the millikelvin temperature range below the critical temperature.
Soon after the discovery of superconductivity in itinerant ferromagnet
UGe$_2$, two new itinerant ferromagnetic superconductors were
discovered---zirconium zinc ZrZn$_2$ and uranium rhodium germanium
URhGe.

\subsubsection{Nickel borocarbides}

The nickel borocarbide class
of superconductors has the general formula $R$Ni$_2$B$_2$C, where $R$
is a rare earth being either magnetic (Tm, Er, Ho, or Dy) or nonmagnetic
(Lu and Y). In the case when $R$ = Pr, Nd, Sm, Gd or Tb in $R$Ni$_2$B$_2$C,
the Ni borocarbides are not superconducting at low temperatures but
antiferromagnetic. In the Ni borocarbides with a magnetic rare earth,
superconductivity coexists at low temperatures with a long-range
antiferromagnetic order. Interestingly, while in the superconducting heavy
fermions with a long-range antiferromagnetic order $T_N \sim 10 T_c$,
in some Ni borocarbides it is just the opposite, $T_c \sim 10 T_N$. Thus,
antiferromagnetism appears deeply in the superconducting state.

Superconductivity in the Ni borocarbides was discovered in 1994 by Eisaki
and co-workers. Transition temperatures in these quaternary intermetallic
compounds can be as high as 17 K.  The Ni borocarbides have a
layered-tetragonal structure
alternating $R$C sheets and Ni$_2$B$_2$ layers. As a consequence, the
superconducting properties of the Ni borocarbides are also anisotropic.
The layered borocarbides DyB$_2$C and HoB$_2$C without Ni also
superconduct, with $T_c =$ 8.5 and 7.1 K, respectively.
Other related compounds, such as the Ni boronitride
La$_3$Ni$_2$B$_2$N$_3$, are also found to superconduct.

\subsubsection{Strontium ruthenate}

Nearly 40 years ago it was found that SrRuO$_3$ is a ferromagnetic
metal with a Curie
temperature of 160 K. In its cousin, Sr$_2$RuO$_4$,
the superconducting state with $T_c \approx$ 1.5 K was discovered
in 1994 by Maeno and his collaborators. The crystal structure of
Sr$_2$RuO$_4$ is layered perovskite, and almost isostructural
to the high-$T_c$ parent compound La$_2$CuO$_4$, in which
the CuO$_2$ layers are substituted by the RuO$_2$ ones.
Recently, bilayer and trilayer strontium ruthenates have been synthesized:
Sr$_3$Ru$_2$O$_7$ is an enhanced paramagnetic metal, and
Sr$_4$Ru$_3$O$_{10}$ is ferromagnetic with a
Curie temperature of 105 K.

\subsubsection{Ruthenocuprates}

Ruthenocuprates are in a sense a hybrid of superconducting cuprates
and the strontium ruthenate. As a consequence, they have a number of
common features with the cuprates, but also many differences. Basically,
there are two ruthenocuprates that superconduct at low temperatures.
The general formulas of these ruthenocuprates are
RuSr$_2 R$Cu$_2$O$_8$ and RuSr$_2 R_2$Cu$_2$O$_{10}$ with
$R$ =  Gd, Eu and Y. The second
ruthenocuprate was discovered first in 1997. The crystal structure of
RuSr$_2 R$Cu$_2$O$_8$ is similar to that of YBa$_2$Cu$_3$O$_7$ except for
the replacement of one-dimensional CuO chains by two-dimensional RuO$_2$
layers. It is assumed that the RuO$_2$ layers act as charge
reservoirs for the CuO$_2$ layers. The principal feature of the
ruthenocuprates is that they are magnetically ordered below
$T_m \sim$ 130 K, and become superconducting at $T_c \sim$ 40 K.
For RuSr$_2 R$Cu$_2$O$_8$, $T_m =$ 130--150 K and $T_c$ = 30--45 K,
while for RuSr$_2 R_2$Cu$_2$O$_{10}$, $T_m =$ 90--180 K and
$T_c$ = 30--40 K. It is believed that the magnetic order arises from
ordering of Ru ions in the RuO$_2$ layers, while the transport occurs in
the CuO$_2$ layers. Superconductivity and the magnetic order are found to
be homogeneous.
There is a consensus that in the ruthenocuprate, there is a small
ferromagnetic component; however, there is no agreement on its origin.
It may originate not only from the Ru moments but also, for example, from
the Gd spins.

\subsubsection{MgCNi$_3$}

Superconductivity in MgCNi$_3$ was discovered in 2001 by Cava and
co-workers, a few months later than that in MgB$_2$. The crystal
structure of MgCNi$_3$ is cubic-perovskite, and similar
to that of BKBO. The perovskite MgCNi$_3$ is special in
that it is neither an oxide nor does it contain any copper. Since Ni is
ferromagnetic, the discovery of superconductivity in MgCNi$_3$ was
surprising. The critical temperature is near 8 K. MgCNi$_3$ is metallic,
and the charge carriers are electrons which are derived predominantly
from Ni.
Structural studies of MgCNi$_3$ reveal structural inhomogeneity.
Apparently, the perovskite cubic structure of
MgCNi$_3$ is modulated locally by the variable stoichiometry on the C sites.

\subsubsection{Cd$_2$Re$_2$O$_7$}

Although Cd$_2$Re$_2$O$_7$ was synthesized in 1965, its
physical properties remained almost unstudied. Unexpectedly,
superconductivity in Cd$_2$Re$_2$O$_7$ was discovered in the
second half of 2001 by Sakai and co-workers. The critical temperature of
Cd$_2$Re$_2$O$_7$ is low, $T_c$ = 1--1.5 K. This compound is the first
superconductor found among the large family of pyrochlore oxides with
the formula A$_2$B$_2$O$_7$, where A is either a rare earth or a late
transition metal, and B is a transition metal. In this structure, the A and B
cations are 4- and 6-coordinated by oxygen anions. The A-O$_4$ tetrahedra
are connected as a pyrochlore lattice with straight A-O-A bonds, while
B-O$_6$ octahedra form a pyrochlore lattice with the bent B-O-B bonds
with an angle of 110--140$^{\circ}$. Assuming that electronic structure
in Cd$_2$Re$_2$O$_7$ as formally Cd$^{2+}$ 4d$^{10}$ and
Re$^{5+}$ 4f$^{14}$5d$^2$, the electronic and magnetic properties are
primarily dominated by the Re 5$d$ electrons.

Oxide superconductors with non-perovskite structure are rare.
Previous studies indicate that the pyrochlores, like the spinels, are
geometrically frustrated. The effect of geometric frustration on the physical
properties of spinel materials is drastic, resulting in, for example,
heavy-fermion
behavior in LiV$_2$O$_4$. Another spinel compound LiTi$_2$O$_4$ is a
superconductor below $T_c =$ 13.7 K. Indeed, x-ray diffraction studies
performed under high pressure showed that superconductivity in
Cd$_2$Re$_2$O$_7$ is detected only for the phases with a structural
distortion. It was suggested that the charge fluctuations of Re ions play
a crucial role in determining the electronic properties of
Cd$_2$Re$_2$O$_7$.

\subsubsection{Na$_x$CoO$_2 \cdot y$H$_2$O}

One of the newest superconductors is the layered cobalt oxyhydrate
Na$_x$CoO$_2 \cdot y$H$_2$O ($\frac{1}{4} < x < \frac{1}{3}$ and
$y =$ 1.3--1.4). The structure of the parent compound
Na$_x$CoO$_2$ consists of alternating layers of CoO$_2$ and Na.
In the hydrated Na$_x$CoO$_2$, the water molecules form
additional layers, intercalating all CoO$_2$ and Na layers. After
the hydration of Na$_x$CoO$_2$, the $c$-axis lattice parameter
increases from 11.16 \AA \, to 19.5 \AA. Thus, the elementary
cell of Na$_x$CoO$_2 \cdot y$H$_2$O consists of three layers of
CoO$_2$, two layer of Na$^+$ ions and four layers of H$_2$O.
Within each CoO$_2$ layer, the Co
ions occupy the sites of a triangular lattice. The 1 - $x$
fraction of Co ions is in the low spin $S = \frac{1}{2}$ Co$^{4+}$
state, while the $x$ fraction is in the $S =$ 0 Co$^{3+}$ state.
In the triangular lattice, the spins of Co$^{4+}$ ions are
ordered antiferromagnetically.

Superconductivity in Na$_x$CoO$_2 \cdot y$H$_2$O occurs in the
CoO$_2$ layers. The superconducting phase as a
function of $x$ has a bell-like shape, situated between 0.25 and
0.33 with a maximum $T_c \simeq$ 4.5 K near $x =$ 0.3.
Na$_x$CoO$_2 \cdot y$H$_2$O is the first superconductor
containing water (ice). All experimental facts indicate that the presence
of water is crucial to superconductivity.

\subsubsection{Hydrides and deuterides}

In addition to the nitrides and carbides from the second group of
superconductors, another class of superconducting compounds that also
have the NaCl structure includes hydrides and deuterides (i.e. compounds
containing hydrogen or deuterium). However, in contrast to the nitrides and
carbides, superconducting hydrides and deuterides are magnetic. In the
seventies it was discovered that some metals and alloys, not being
superconducting in pure form, become relatively good superconductors when
they form alloys or compounds with hydrogen or deuterium. These metals
include the transition elements palladium (Pd) and thorium (Th) that have
unoccupied 4$d$- and 5$f$-electron shells, respectively.

In 1972, Skoskewitz discovered that the transition element Pd which has a
small magnetic moment normally preventing the pairing of electrons, joins
hydrogen and forms the PdH compound that superconducts at $T_c$ = 9 K.
Later on, it was found that by doping such a system with noble metals the
critical temperature increases up to 17 K. Interestingly, the
palladium-deuterium compound also superconducts, and its critical
temperature equal to 11 K is higher than that of PdH. So the hydrogen
isotope effect in PdH is reverse (negative). In contrast, the critical
temperatures of the ThH and ThD compounds do not differ drastically from
each other like those of PdH and PdD.

\section{Physical properties of the third group of \\superconductors}

In the previous section, superconducting materials were classified into
three groups. The first group consists of conventional superconductors; the
second comprises half-conventional ones, and unconventional superconductors
form the third group. Assume that one day a room-temperature superconductor
will become available. What group will it belong to? The answer is more or less
obvious: to the third group of unconventional superconductors. Indeed,
the highest critical temperature in third group of superconductors,
$T_{c,max} \simeq$ 135 K, is more than three times higher than that of the
second group and more than one order of magnitude higher than that
of group of conventional superconductors. Moreover, it seems that the critical
temperatures of superconductors of the first and the second groups cannot
exceed 10 K and 40 K, respectively. The third group of superconductors is the
largest and has the highest rate of growth in the last twenty five years.
If the rate of the growth of the third group will remain in the future at the same
level, then, one will soon need to make an internal classification of this group.

From the discussion in the previous paragraph, it is evident that, in order to
have guide to the properties of room-temperature superconductors, it is
necessary to analyze the physical properties of superconductors of the third
group, and not those of the first and second groups. One may argue, however,
that room-temperature superconductors can form a separate group of
superconductors, i.e. the fourth one. Generally speaking such a situation may
occur, but there are only two options for a low-dimensional compound to be
either magnetic or non-magnetic. As a consequence, every low-dimensional
superconductor belongs either to the second or third group of superconductors.

Let us enumerate the common physical properties of superconductors of the
third group. All superconductors of the third group:
\begin{itemize}
\item
are magnetic or, at least, have strong magnetic correlations,

\item
are low-dimensional,

\item
have strongly correlated electrons/holes,

\item
are near a metal-insulator transition,

\item
are apparently near a quantum critical point,

\item
are type-II superconductors,

\item
have small-size Cooper pairs,

\item
have a low density of charge carriers, $n_s$,

\item
have a universal $n_s \propto \sigma (T_c) \dot T_c$ dependence \cite{7},
where $\sigma(T_c)$ is the {\em dc} conductivity just above $T_c$,

\item
have large values of $H_{c2}$ and $\lambda$ (magnetic penetration
depth) and a large gap ratio $2\Delta_p / (k_B T_c)$, where $\Delta_p$ is
the {\em pairing} energy gap (see Section 6),

\item
have anisotropic transport and magnetic properties,

\item
have a complex phase diagram (if there is a parameter to vary),

\item
have the moderately strong electron-phonon interaction,

\item
have an unstable lattice,

\item
have charge-donor and charge-acceptor sites (i.e. there is a charge transfer), and

\item
have a complex structure (with the exception of hydrides, deuterides
and a few heavy fermions).
\end{itemize}

Among superconductors of the third group:
\begin{itemize}
\item
the $T_c$ value correlates with the behavior of spin fluctuations---the more
dynamic the fluctuations are, the higher the $T_c$ value is,

\item
the localized states in {\em undoped} material have spin-1/2 ground states,

\item
the average $T_c$ value of layered compounds is higher than that of
one-dimensional ones (even so, the transport properties of the layered
compounds are quasi-one-dimensional),

\item
superconductors with $T_c >$ 20 K have no
metal-metal bonds (only heavy fermions have metal-metal bonds),  and

\item
oxides and organic superconductors represent the absolute majority of the
group.
\end{itemize}

At the same time, there are differences among superconductors of the third
group. The two main differences are:
\begin{itemize}
\item
the $T_c$ value of hole-doped superconductors is on average a few times
higher than that of electron-doped superconductors,

\item
the average $T_c$ value of antiferromagnetic compounds is at least one order
of magnitude higher than that of ferromagnetic superconductors.
\end{itemize}

Considering the common features of superconductors of the third group, one
should realize that some of these features are direct consequences of
the others. For example, the anisotropic character of transport and magnetic
properties of superconductors of the third group is a direct consequence of a
low-dimensional structure of these superconductors. The expression ``systems
with strongly correlated electrons'' partially assumes that the electron-phonon
interaction in these systems is sufficiently strong, and they have a complex
phase diagram. The moderately strong electron-phonon interaction results in a
large value of the pairing energy gap and, therefore, in a large value of the gap
ratio $2\Delta_p / (k_B T_c)$. Since in
superconductors of the third group, the Cooper pairs have a small size and a low
density, this leads to the penetration depth and, consequently, the ratio
$\lambda / \xi$ to be large, where $\xi$ is the coherence length. Therefore, all
superconductors of the third group are type-II. The so-called
Homes law, $n_s \propto \sigma (T_c) \dot T_c$, \cite{7} literally means that an
high-temperature superconductor should be a bad conductor just above $T_c$.
The same conclusion follows from the other experimental fact that
superconductivity with an high $T_c$ occurs near a metal-insulator transition.
Hence, some of these common features of superconductors of the third group
are more important than others. Let us select the most important ones.

\section{Requirements for high-$T_c$ materials}

We are now in a position to discuss the most important requirements for
materials that superconduct at high temperatures. From the previous section,
one can conclude that a room-temperature superconductor should:
\begin{itemize}
\item
be hole-doped,

\item
be low-dimensional (preferentially, layered but with quasi-one-dimensional
transport properties),

\item
be antiferromagnetic or, at least, have strong magnetic correlations,

\item
have strongly correlated holes,

\item
be near a metal-insulator Mott transition,

\item
have an {\em undoped} parent compound with
the spin-1/2 localized states,

\item
have {\bf dynamic} spin fluctuations,

\item
be in a state above a quantum critical point,

\item
have an unstable lattice,

\item
have a complex structure (i.e. with more than two sites per unit cell),

\item
have electron-acceptor sites, and

\item
have no metal-metal bonds.
\end{itemize}

It is necessary to comment on the issue of quantum critical point.
In a quantum critical point where a magnetic order is about to form or to
disappear, the spin fluctuations are the strongest. At the moment of writing,
we do not know yet how to determine from a single measurement the
presence/absence of a quantum critical point in a certain compound. So, this
can be the first intermediate goal: how to determine quickly the
presence/absence of a quantum critical point in a given compound.

In addition to these common features of the third group of superconductors,
consider one more observation concerning the structure of good superconductors.
In cuprates, the unit cell has three {\em interacting} subsystems: a quasi-metallic,
a magnetic ones and charge reservoirs. The charge reservoirs in cuprates are the
layers that intercalate the CuO$_2$ layers. After accepting/donating electrons,
the charge reservoirs become semiconducting or insulating. The quasi-metallic
and magnetic subsystems in cuprates are located into the CuO$_2$ planes,
resulting in the phase separation. In organic superconductors, however, the second
and the third subsystems coincide: the charge reservoirs, after donating/accepting
electrons to/from organic molecules, become magnetic. For instance, the structure
of a Bechgaard salt shown in Fig. 2 is a good example. Thus, in organic
superconductors, one subsystem performs two functions. To summarize, a
potential room-temperature superconductor should have:

1) {\bf a quasi-metallic subsystem},

2) {\bf charge reservoirs}, and

3) {\bf magnetic atoms/molecules}.

\noindent
Each subsystem should be coupled to the two others. Experimentally, the
second and the third subsystems can be represented by the same
atoms/molecules.

The requirements summarized in these section basically include Geballe's
guidelines, described in Section 2. We shall use these hints in Sections 7--9.

\section{Principles of superconductivity}

Considering magnetic and structural requirements for materials, we should
remember that the materials must superconduct after all.
What does superconductivity as a phenomenon require?
Let us discuss in this section the main principles of
superconductivity as a phenomenon, valid for every superconductor
independently of its characteristic properties and material. The underlying
mechanisms of superconductivity can be different in various
materials, but certain principles must be satisfied. One should however
realize that the principles of superconductivity are not limited to those
discussed in this section: it is possible that there are others which we
do not know yet about. More detailed description of the principles of
superconductivity can be found elsewhere \cite{2,3}.

The first principle of superconductivity is:

\vspace{0.5cm}
\noindent
Principle 1: \hspace*{3mm}
{\bf Superconductivity requires quasiparticle pairing}
\vspace{0.5cm}

The electron (hole) pairs are known as Cooper pairs.
In solids, superconductivity as a quantum state cannot occur without the
presence of bosons. Fermions are not suitable for forming a quantum state
since they have spin and, therefore, they obey the Pauli exclusion principle
according to which two identical fermions cannot occupy the same quantum
state. Electrons are fermions with a spin of 1/2, while the Cooper
pairs are already composite bosons since the value of their total spin is
either 0 or 1. Therefore, the electron (hole) pairing is an inseparable part of
the phenomenon of superconductivity and, in any material, superconductivity
cannot occur without quasiparticle pairing.

In the framework of the BCS theory [5], the electron pairing occurs
in momentum space. However, for the occurrence of superconductivity in the
general case, the electron pairing may take place not only in momentum  space
but also in real space. The electron pairing in momentum space can be considered
as a {\em collective} phenomenon, while that in real space as {\em individual}.
Independently of the space where they are paired---momentum or real---two
electrons can form a bound state {\bf only if} {\em the net force acting between
them is attractive}.

Superconductivity requires the electron pairing and the Cooper-pair
condensation. The second principle of superconductivity
deals with the Cooper-pair condensation taking place at $T_c$. This
process is also known as the onset of long-range phase coherence.

\vspace{0.5cm}
\noindent
Principle 2:\hspace*{3mm} {\bf The transition into the superconducting
state is the Bose-\\ \hspace*{24mm}Einstein-like condensation and occurs in
momentum space}
\vspace{0.5cm}

The two processes---the electron pairing and the onset of phase coherence---are
independent of one another. Superconductivity requires both. In conventional
superconductors, the pairing and the onset of phase coherence take place
simultaneously at $T_c$. In many unconventional superconductors, quasiparticles
become paired above $T_c$ and start forming the superconducting condensate
only at $T_c$.

In the 1920s, Einstein predicted that if an ideal gas of identical atoms,
i.e. bosons, at thermal equilibrium is trapped in a box, at sufficiently
low temperatures the particles can in principle accumulate in the lowest
energy level. This may take place only if the quantum wave packets of
the particles overlap. In other words, the wavelengths of the matter waves
associated with the
particles---the {\em Broglie waves}---become similar
in size to the mean particle distances in the box. If this happens, the
particles condense, almost motionless, into the lowest quantum state,
forming a Bose-Einstein condensate.
The two condensates---superconducting and Bose-Einstein---have common
quantum properties, but also, they have a few differences, which have
been discussed elsewhere \cite{2}.

The third principle of superconductivity is:

\vspace{0.5cm}
\noindent
Principle 3:\hspace*{3mm} {\bf The mechanism of electron
pairing and the mechanism of\\ \hspace*{24mm}Cooper-pair condensation
must be different}
\vspace{0.5cm}

The validity of the third principle of superconductivity will be evident
after the presentation of the fourth principle. Historically, this principle
was introduced first. To recall, in conventional superconductors, phonons
mediate the electron pairing, while the overlap of wavefunctions ensures
the Cooper-pair condensation. Generally speaking, if in a superconductor,
the same ``mediator'' (for  example, phonons) is responsible for the
electron pairing and for the  onset of long-range phase coherence
(Cooper-pair condensation), this will lead to the  collapse of superconductivity.

If the first three principles of superconductivity do not deal with
numbers, the forth principle can be used for making various estimations.

Generally speaking, a superconductor is characterized by a pairing energy
gap $\Delta_p$ and a phase-coherence gap $\Delta_c$.
For genuine (not proximity-induced) superconductivity, the
phase-coherence gap is proportional to $T_c$:
\begin{equation}
2\Delta_c = \Lambda\,k_B T_c,
\end{equation}
where $\Lambda$ is the coefficient proportionality,
and $k_B$ is the Boltzmann constant.
At the same time, the pairing energy gap is proportional
to the pairing temperature $T_{pair}$:
\begin{equation}
2\Delta_p = \Lambda ' \,k_B T_{pair}.
\end{equation}
Since the formation of Cooper pairs must precede the onset of
long-range phase coherence, then in the general case,
$T_{pair} \geq T_c$.

In conventional superconductors, however, there is only one energy gap
$\Delta$ which is in fact a pairing gap but proportional to $T_c$:
\begin{equation}
2\Delta = \Lambda\,k_B T_c,
\end{equation}
This is because, in conventional superconductors, the electron pairing and
the onset of long-range phase coherence take place at the same
temperature---at $T_c$. In all known cases, the coefficients $\Lambda$
and $\Lambda '$ lie in the interval between 3.2 and 6 (in one heavy
fermion, $\sim$ 9) \cite{2,3}. We are now in position to discuss
the fourth principle of superconductivity:

\vspace{0.5cm}
\noindent
Principle 4:\hspace*{1mm} {\bf For genuine, homogeneous
superconductivity, $\Delta_p > \Delta_c > \frac{3}{4}k_BT_c$
\\ \hspace*{22mm}always (in conventional superconductors,
$\Delta > \frac{3}{4}k_BT_c$)
}
\vspace{0.5cm}

The reason why superconductivity occurs exclusively at low temperatures
is the presence of substantial thermal fluctuations at high temperatures.
In conventional
superconductors, the energy of electron binding, $2\Delta$, must be
larger than the thermal energy; otherwise, the pairs will be broken up
by thermal fluctuations.
In the case of unconventional superconductors, the same reasoning is also
applicable for the phase-coherence energy gap, $\Delta_c$.
The last inequality, $\Delta_p > \Delta_c$ for unconventional superconductors,
means that the pairing energy $2\Delta_p$ of the Cooper pairs must be larger
than the strength of the coupling of bosonic excitations responsible for
mediating the long-range phase coherence with the Cooper pairs, which is
measured by the energy $2\Delta_c$. If the strength of this coupling
will exceed the pairing energy $2\Delta_p$, the Cooper pairs will
immediately be broken up.

In the case $\Delta_p = \Delta_c$ occurring at some temperature $T < T_c$
remaining constant, locally there will be superconducting fluctuations due
to thermal fluctuations, thus, a kind of inhomogeneous superconductivity.
It is necessary to mention that the case $\Delta_p = \Delta_c$ must not
be confused with the case $T_{pair} = T_c$, because usually
$\Lambda ' > \Lambda$.

Finally, let us go back to the third principle of superconductivity to show
its validity. The case in which the same bosonic excitations mediate the
electron pairing {\bf and} the phase coherence is equivalent to the case
$\Delta_p = \Delta_c$ discussed above. Since, in this particular case, the
equality $\Delta_p = \Delta_c$ is independent of temperature, the
occurrence of homogeneous superconductivity is impossible.

\section{An approach to room-temperature superconductivity}

Approaching the problem of room-temperature superconductivity, different
resear\-chers can use the information presented above in different ways:
human imagination does not have limits. One approach, however, is more
or less obvious. The structure of cuprates basically satisfies all the requirements
described above. Then, one can replace the CuO$_2$ planes by planes of different
type, which can accommodate, depending on the doping level, the Cooper pairs
and antiferromagnetic ordering above room temperature. The only problem is
that one should know what type of planes to use.

Another possibility is to use the structure of organic salts shown in Fig. 2 as the
basis for the structure of new superconductors. Luckily, the issue of the
presence of Cooper pairs in some organic compounds above room temperature is
known already for some time. In addition to substitution of the organic molecules,
the magnetic molecules of PF$_6$ in Fig. 2 should be replaced by other type of
molecules which are ordered antiferromagnetically above room temperature.
To the end of this chapter, we shall explore this approach to the problem.

\section{Organic molecules, polymers and tissues with \\electron pairs
above room temperature}

According to the first principle of superconductivity, superconductivity requires
electron pairing. Indeed, the electron pairing is the keystone of superconductivity.
Therefore, in quest of compounds that superconduct above room temperature,
one should first look at materials which tolerate the presence of Cooper pairs at
high temperatures.
Fortunately, it is known already for some time that the Cooper pairs exist in some
organic compounds at and above room temperature. In this section we are going to
discuss these organic materials. It is worth to mention, however, that
superconductivity does not occur in these compounds because superconductivity
requires not only electron pairing but also the establishment of long-range
phase coherence.

The idea to use organic compounds as superconducting materials is not new. In
1964, Little proposed a model in which a high $T_c$ is obtained due to a non-phonon
mediated mechanism of electron attraction, namely, an {\em exciton} model for
Cooper-pair formation in long chainlike organic molecules \cite{8}. In the framework
of his model, the maximum critical temperature was estimated to be around 2200 K!
Little's paper has encouraged the search for room-temperature superconductivity,
especially in organic compounds.

\subsection{Organic molecules}

In 1975, Kresin and co-workers showed that the
superconducting-like state exists {\em locally} in complex organic molecules
with conjugate bonds \cite{9}. Figure 3 shows a few examples of such molecules.
\begin{figure}[t]
\epsfxsize=0.8\columnwidth
\centerline{\epsffile{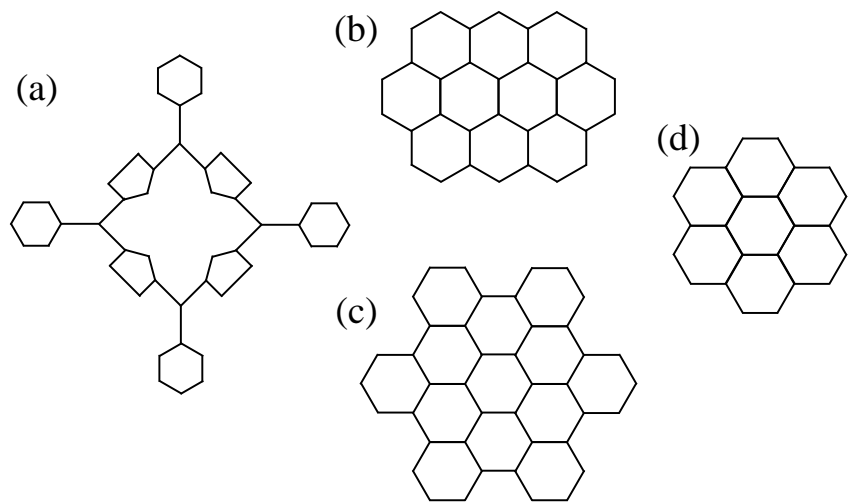}}
\caption{Organic molecules with delocalized $\pi$ electrons:
(a) tetraphenylporphin; (b) ovalene; (c) hexabenzocoronene, and (d) coronene
\protect\cite{9}.}
\end{figure}
Their main building blocks are carbon and
hydrogen atoms. The characteristic feature of these conjugated hydrocarbons
is the presence of a large number of $\pi$ electrons. These collectivized
electrons are in the field of the so-called $\sigma$ electrons which are
located close to the atomic nuclei and not much different from the ordinary
atomic electrons. At the same time, the $\pi$ electrons are not localized
near any particular atom, and they can travel throughout the entire
molecular frame. This makes the molecule very similar to a metal. The
framework of atoms plays the role of a crystal lattice, while the $\pi$
electrons that of the conduction electrons. It turns out, in fact, that the
conjugated hydrocarbons with even number of carbon atoms are more than
just similar to a metal, but are actually small superconductors \cite{9}.
Experimentally, conjugated hydrocarbons with even number of carbon
atoms (thus, with even number of $\pi$ electrons) exhibit properties
similar to those of a superconductor: the Meissner-like effect, zero
resistivity and the presence of an energy gap. The $\pi$ electrons form
bound pairs analogous to the Cooper pairs in an ordinary superconductor.
The pair correlation mechanism is principally due to two effects: (i) the
polarization of the $\sigma$ electrons, and (ii) $\sigma - \pi$ virtual
electron transitions. However, if the number of $\pi$ electrons is odd, the
properties of such conjugated hydrocarbons are different from those of a
superconductor.

$\bullet$ The alkali-doped fullerenes discussed in Section 3 are able to
superconduct but they are electron-doped. To exhibit
room-temperature superconductivity, the single crystals of C$_{60}$ must be
doped by holes. Thus, one should find suitable dopant species for this purpose.
Theoretical calculations show that fullerenes having a diameter smaller than that
of C$_{60}$, such as C$_{28}$ and C$_{20}$, are able to exhibit a higher value
of $T_c$ relative to that of C$_{60}$ \cite{2}. Unlike graphene and other long
organic polymers, the fullerenes have an advantage to be packed into any form.

One can use fullerenes not only in pure but also in polymerized form. As an
example, Figure 4 shows various one- and two-dimensional polymeric solids
formed from C$_{60}$.
\begin{figure}[t]
\epsfxsize=0.75\columnwidth
\centerline{\epsffile{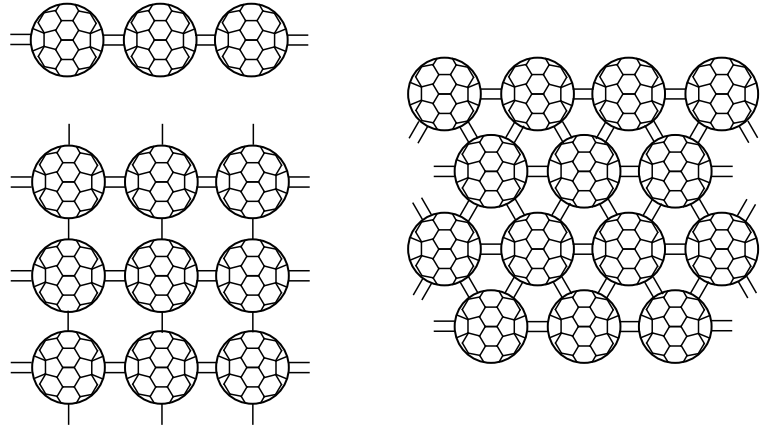}}
\caption{Various one- and two-dimensional polymeric solids
formed from C$_{60}$ \protect\cite{10}. The C$_{60}$ balls are shown
schematically.}
\end{figure}

$\bullet$ As discussed in Section 3, carbon nanotubes can be viewed as giant
conjugated molecules with a conjugated length corresponding to the whole
length of the tube. The nanotubes are also a promising candidate with which to
form a room-temperature superconductor.

The single-walled carbon nanotubes with a diameter of 4.2 $\pm$ 0.2 \AA \,
exhibit {\em bulk} superconductivity below $T_c \simeq$ 15 K. The
nanotubes with a smaller diameter may display a higher $T_c$. The onset
of {\em local} superconductivity at 645 K was observed in single-walled carbon
nanotubes containing a small amount of the magnetic impurities Ni\,:\,Co
($\leq$ 1.3 \%) \cite{11}. This unconfirmed evidence is based on transport,
magnetoresistance, tunneling and Raman measurements. In single-walled
carbon nanotubes, the energy gap obtained in tunneling measurements is about
$\Delta_p \simeq$ 100 meV \cite{11}. By embedding these nanotubes into a
{\em dynamic} magnetic medium, one can witness bulk superconductivity above
450 K. Thus, in the general case, in order to accommodate the pairs at
high temperatures, nanotubes should be {\em doped} or contain {\em defects}.

\subsection{Organic conjugated polymers}

Polythiophene is a one-dimensional conjugated polymer. Figure 5(a) shows its
structure. It has been known already for some time that, in polythiophene, the
dominant nonlinear excitations are positively-charged polarons and bipolarons
\cite{12}. This means that the Cooper pairs with a charge of $+2|e|$ exist at room
temperature in polythiophene. Figure 5(b) depicts a schematic structural diagram
of a bipolaron on a polythiophene chain.
In a thiophene ring, the four carbon $p$ electrons and the two sulfur $p$ electrons
provide the six $p$ electrons that satisfy the $(4n\,+\,2)$ condition necessary
for aromatic stabilization. Polythiophene has a few derivatives and one of them
shown in Fig. 5(c) is called poly(3-alkylthienylenes) or P3AT for short. In contrast
to polythiophene, P3AT is soluble.
\begin{figure}[t]
\epsfxsize=0.6\columnwidth
\centerline{\epsffile{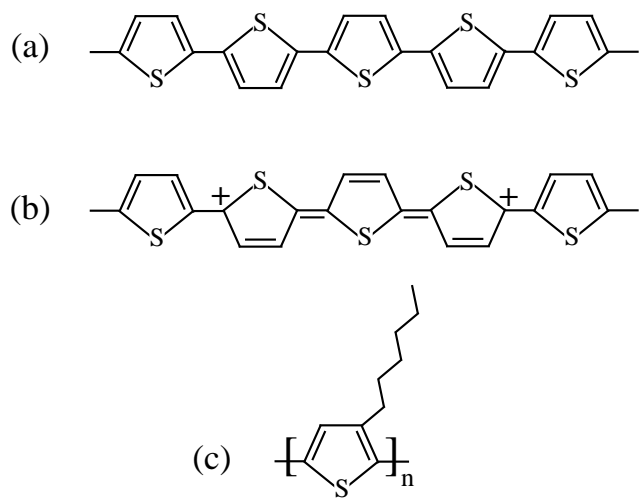}}
\caption{(a) Chemical structure of polythiophene.
(b) Schematic structural diagram of a positively-charged polaron
on a polythiophene chain. (c) Chemical structure of soluble
poly(3-alkylthienylenes) \protect\cite{12}.}
\end{figure}

Polythiophene chains have the infinite length. Polythiophene chains having a finite
length can be used too. However, the ends of the pieces of the polymer must be
``closed.'' Alternatively, the two ends of a polymer piece can be attached one
to another leading to the formation of a ring. Taking into account that the width of
a bipolaron is a few lattice constants; then, the length of pieces of polythiophene
chains, $\ell$, should be 2--3 times larger; thus $\ell \sim 15a$, where $a$
is the lattice constant.

Other conjugated polymers also contain bipolarons. It is known that
positively-charged bipolarons exist, for example, in polyparaphenylene, polypyrrole
and poly(2,5-diheptyl-1,4-phenylene-alt-2,5-thienylene) (PDHPT) \cite{12}. The
structure of polyparaphenylene is depicted in Fig. 6(a).  A bipolaron on a
polyparaphenylene chain is schematically shown in Fig. 6(b). One of the
derivatives of polyparaphenylene, $p$-sexiphenyl depicted in Fig. 6(c). The
structure of PDHPT is illustrated in Fig. 6(d). Bipolarons have also been observed
in other one-dimensional conjugated organic polymers such as polyparaphenylene
and polypyrrole \cite{12}. A large number of conjugated
polymers used in electroluminescent diodes at room temperature contain
bipolarons, such as poly($p$-phenylene vinylene) etc., (see Fig. 2 in \cite{13}).
All these polymers are commercially available.
\begin{figure}[t]
\epsfxsize=0.65\columnwidth
\centerline{\epsffile{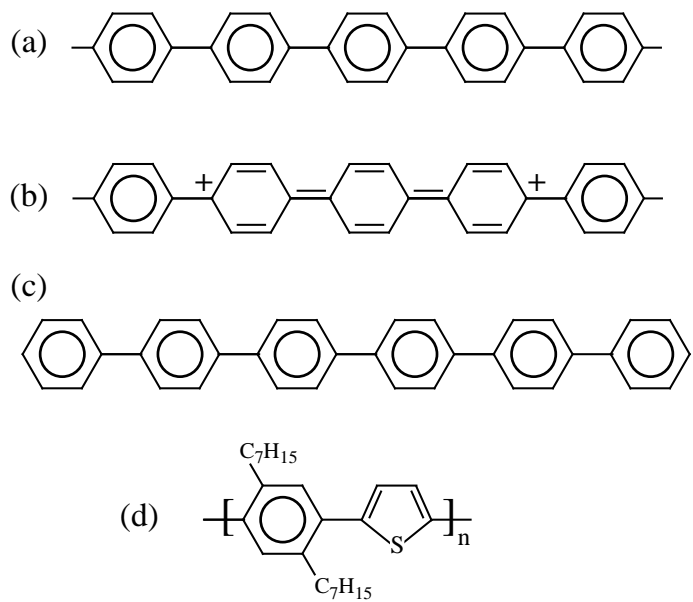}}
\caption{(a) Chemical structure of polyparaphenylene.
(b) Schematic structural diagram of a positively-charged bisoliton
on a polyparaphenylene chain \protect\cite{12}. (c) Molecular structure of
$p$-sexiphenyl, and (d) the chemical structure of soluble
poly(2,5-diheptyl-1,4-phenylene-alt-2,5-thienylene) (PDHPT).}
\end{figure}

\subsection{Living tissues}

It is known that (i) in redox reactions occurring in living organisms,
electrons are transferred from one molecule to another in pairs with
opposite spins; and (ii) electron transport in the synthesis process of ATP
(adenosine triphosphate) molecules in conjugate membranes of mitochondria
and chloroplasts is realized by pairs, not individually \cite{14,15,16}.

In living tissues, the electron pairing simplifies their propagation because the
calculations show that, for electrons, it is more profitable {\it energetically} to
propagate together than separately, one by one \cite{15}. So, the electron
pairing occurs
in living tissues first of all because of an energy gain; the electron spin is a
secondary reason for the pairing. However, we are more interested in electron
pairing because of spin. The body temperature of living creatures is usually near
the room temperature. Therefore, one can use these materials to form a
room-temperature superconductor. As discussed earlier,
superconductivity does not occur in living tissues because, according to the
principles of superconductivity, it requires not only the electron pairing but also
the onset of long-range phase coherence.

Recently, living organisms have been found in extreme conditions: some survive
without sunlight, some survive in water near the boiling point. Probably the most
extreme case is the discovery of so-called black smokers or chimneys on the
ocean floor. Deep-sea vents provide an unusual habitat for some primitive forms
of extremophile bacteria and deep-sea crabs that can survive extreme conditions.
For example, the {\it Spire} vent is located at the Broken Spur Vent Field in the
Mid-Atlantic Ridge, 3080 m below sea level. The measurement of water
temperature at which the crabs reside there yielded $T \sim$ 365 C. This means
that the electron pairs exist in certain organic materials at temperatures above
600 K.

Mitochondria and chloroplasts are integral parts of almost every living cell. In
principle, one can easily use their membranes to form a room-temperature
superconductor. The redox reactions occur practically in every cell. One should
find out what parts of the cells are responsible for the redox reactions, and then
use these tissues. The living tissues are usually one- or two-dimensional.

Other living tissues may also contain bipolarons. Nowadays it is known that, in
the living matter, the signal transfer occurs due to charge (electron) transfer. It
is possible that, in {\em some} cases, the electron transfer occurs in pairs with
opposite spins.

Pullman and Pullman already in 1963 emphasized that ``the essential fluidity of
life agrees with the fluidity of the electronic cloud in conjugated molecules. Such
systems may thus be considered as both the cradle and the main backbone of life''
\cite{17}. Indeed, all natural molecules are conjugated \cite{18}, and it is possible
that, some of them can be used to synthesize a room temperature superconductor.

\subsubsection{Graphite}

For the last forty years, graphite is one of the most studied materials
(see, e.g. \cite{19}). It is also one of the {\bf most} promising superconducting
materials. Graphite intercalation compounds (GICs) able to superconduct were
discussed in Section 3. All superconducting GICs are alkali-doped and,
therefore, magnetic due to alkali spins ordered antiferromagnetically. In the
superconducting GICs, the charge carriers are however electrons, not holes.

There exist both theoretical predictions and experimental evidence that
electronic instabilities in pure graphite can lead to the occurrence of
superconductivity and ferromagnetism, even at room temperature \cite{20}.
In graphite, an intrinsic origin of high-temperature superconductivity relates to
a {\bf topological disorder} in graphene layers \cite{20}. This disorder enhances
the density of states at the Fermi level. For example, four hexagons in graphene
can in principle be replaced by two pentagons and two heptagons. Such a defect
in graphene modifies its band structure. In real space, the disorder in graphene
transforms an ideal two-dimensional layer into a network of
quasi-one-dimensional channels.

In practice, the graphene sheets are always finite. Their electronic properties
are drastically different from those of bulk graphite. It is experimentally
established that the electronic properties of nanometer-scale graphite are
strongly affected by the structure of its edges \cite{20}. The graphene edges
induce electronic states near the Fermi level. Any graphene edge can be presented
by a linear combination of the two basic edges: zigzag and armchair, shown in
Fig. 7. The free energy of an armchair edge is lower than that of a zigzag edge.
\begin{figure}[t]
\epsfxsize=0.3\columnwidth
\centerline{\epsffile{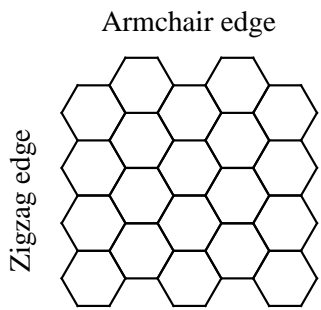}}
\caption{Two basic types of graphite edges.}
\end{figure}

The appearance of bipolarons at high temperatures in graphene depends on the
graphene structure: graphene sheets must be topologically disordered. As
mentioned above, one can replace some hexagons in graphene by pentagons and
heptagons. Alternatively, some carbon atoms in graphene can be substituted for
B, N or Al (B and Al are probably better than N because each N adds an additional
electron to graphene, while room-temperature superconductivity requires holes).
Instead of graphene sheets, one can for instance use nanostripes of graphene.
Since the bonds between adjacent layers in graphite are weak, individual atomic
graphene planes can be pulled out of bulk crystals \cite{21} and be used further.

In principle, it should be not a problem for bipolarons to occur in graphite and
graphene-based compounds above room temperature. The main problem for the
occurrence of bulk superconductivity in graphite and other organic compounds
above room temperature is the onset of long-range phase
coherence. One can dope graphite by atoms/molecules, spins of which are
ordered antiferromagnetically after the diffusion. This issue will be discussed
in the following section.

At the end, it is worth noting that materials which contain bipolarons above
room temperature are not limited to those discussed in this section. It is
possible that there are compounds which tolerate the presence of bipolarons
above room temperature but are unknown to us at the moment of the writing.

\section{Structure and constituents of a potential \\room-temperature
superconductor}

As discussed in Section 7, the structure of a Bechgaard salt shown in Fig. 2
basically satisfies all the requirements for materials that superconduct at
high temperatures, which are described in Section 5. In superconducting organic
salts, the Cooper pairs reside on organic molecules organized in stacks, which
are shown in Fig. 1. The stacks form one-dimensional structures similar
to those in Fig. 2. In the crystal, the chains of other atoms or molecules
(e.g. PF$_6$, ClO$_4$ etc.) are situated between the stacks and aligned parallel
to them, as shown in Fig. 2. The organic molecules donate electrons to the
anions, which prefer to order antiferromagnetically. In some organic salts, e.g.
in the BEDT-TTF family, the arrays of organic molecules form conducting layers
separated by insulating anion sheets. So, in contrast to the Bechgaard salts
which exhibit quasi-one-dimensional electron transport, the electronic structure
of the BEDT-TTF family is two-dimensional.

As suggested in Section 7, the idea to create a compound able to superconduct
above room temperature is straightforward. Taking the crystal structure of the
Bechgaard salt shown in Fig. 2 as a basis, one should replace the stacks of the
organic TMTSF molecules by molecules/polymers/layers of a certain material
which contains the Cooper pairs above room temperature. At the same time, the
molecules of PF$_6$ in Fig. 2 should be substituted for other atoms or molecules
which are able to accept electrons from the conducting counterparts and to
become antiferromagnetically ordered above room temperature. Also, one can
add a small amount of atoms/molecules which stand duty {\em exclusively} as
charge reservoirs. So, the basic idea is more or less obvious; the main question
is what materials to use and how to achieve the right intercalation.

Materials which contain bipolarons above room temperature have been discussed
in the previous section. As a matter of fact, they are all organic. What is about
atoms/molecules which are magnetic in the intercalated state?

A superconductor of the third group must be magnetic or, at least, have strong
magnetic correlations. While oxides can be magnetic naturally, like cuprates
for example, organic and living-tissue-based compounds must be doped by
magnetic species. Unfortunately, during evolution, Nature did not need to develop
such magnetic materials. Hence, one should only rely on accumulated scientific
experience and work by trial and error.

By doping organic materials or living tissues, one should take into account that,
after the diffusion, the dopant species must not be situated too close to the
organic molecules/tissues. Otherwise, they will have a strong influence on
bipolaron wavefunctions and may even break up the bipolarons. On the other hand,
the dopant species cannot be situated too far from the organic molecules/tissues
because, as the common logic suggests, bipolarons should be coupled to spin
fluctuations. Therefore, synthesizing a room-temperature superconductor, one
must pay attention to its structure: the ``distance'' between failure and success
can be as small as 0.01 \AA \, in the lattice constant.

What materials can be used as the acceptors of electrons?
From experience \cite{2}, materials able to accept
electrons from organic molecules are the following atoms and molecules: Cs, I, Br,
PF$_6$, ClO$_4$, FeCl$_4$, Cu(NCS)$_2$, Cu[N(CN)$_2]$Br and Cu[N(CN)$_2$]Cl.
However, in existing superconducting organic salts, they are ordered
antiferromagnetically at low temperatures, and it is not obvious at all, if they are
able to behave in the same way above room temperature. As discussed in
Section 5, in a superconductor of the third group, spin fluctuations should be
dynamic. As a matter of fact, the dynamic character of spin fluctuations can
be achieved artificially, as suggested elsewhere \cite{2}.

At the end of this section, a few remarks about the structure of a potential
room-temperature superconductor should be made. As an example, consider
polythiophene. Polythiophene, its derivatives and other organic conjugated
polymers are usually doped by using the so-called electrochemical method
\cite{12}. The reaction is carried out at room temperature in an electrochemical
cell with the polymer as one electrode. To remove electrons from organic polymers,
oxidation is usually used. Through doping, one can control the chemical potential.
In practice, however, it is impossible to foresee the structure of a doped organic
compound, even knowing materials before the beginning of the doping procedure.
Depending on their origin, concentration and size, the dopant species after the
diffusion can take different positions relative to the polythiophene chains.
Figure 8 shows several examples of possible positions of dopant species relative
to polythiophene chains. Upon doping, the polythiophene chains and dopant species
can, for example, be in-plane, as sketched in Fig. 8(a), or alternate, as illustrated
in Fig. 8(b). They can, for example, form a checker-board pattern shown
schematically in Fig.  8(c). For instance, in Na-doped polyacetylene, the Na$^+$
ions and polyacetylene chains form a modulated lattice with a ``triangular''
pattern depicted in Fig. 8(d). In Na-doped polyacetylene, such a lattice appears
exclusively at moderate doping levels.
\begin{figure}[t]
\epsfxsize=0.9\columnwidth
\centerline{\epsffile{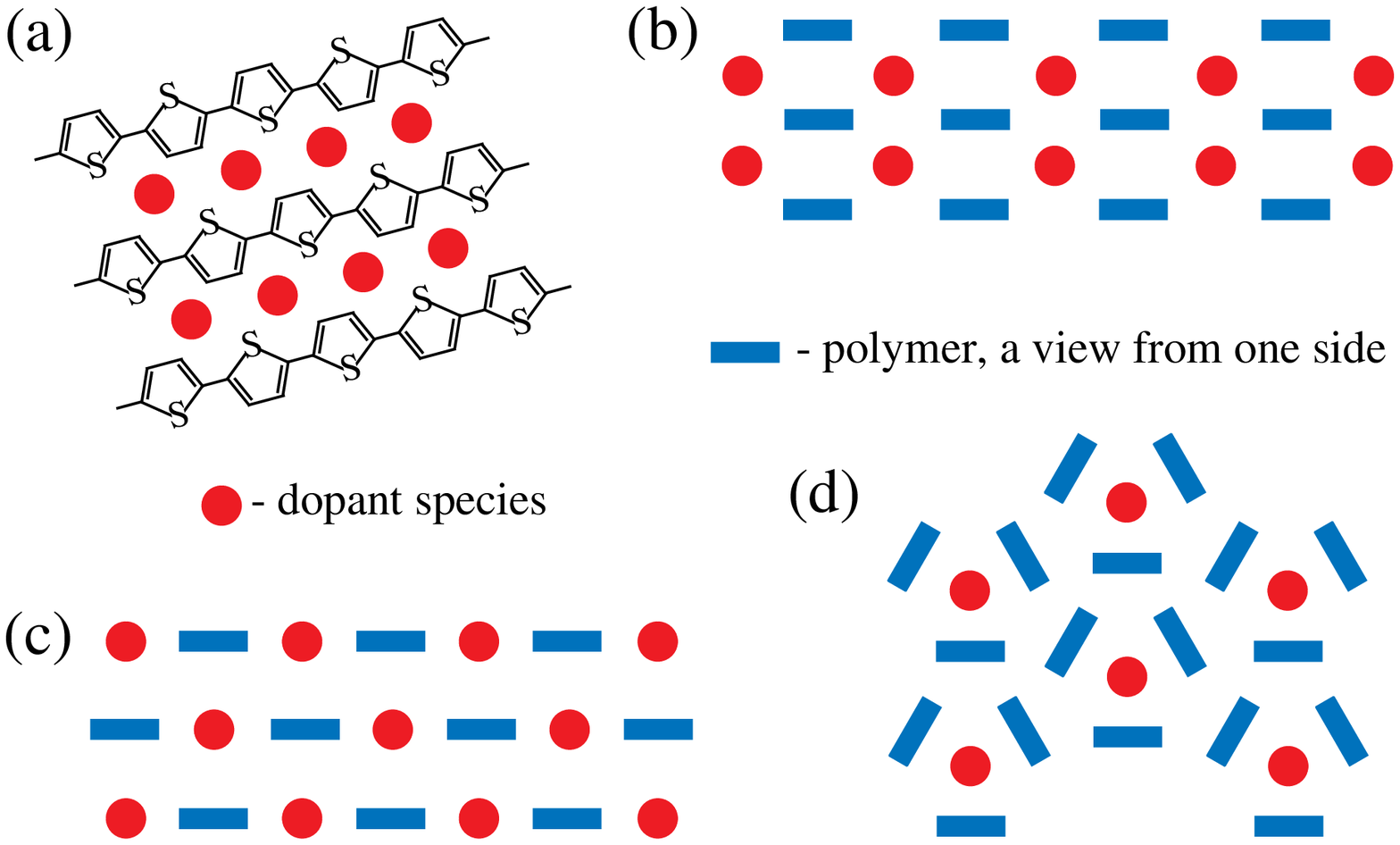}}
\caption{Possible positions of dopant species relative to infinite
polythiophene chains: (a) in the plane of polythiophene chains;
(b) between the planes; (c) a checker-border pattern, and
(d) a ``triangular'' pattern realized in Na-doped polyacetylene \protect\cite{12}.}
\end{figure}

Using various magnetic atoms/molecules, one should not forget to control the
doping level of the organic polymers. It can be done, for example, by adding a
small amount of atoms/molecules of another type, which may or may not be
magnetic after the diffusion. Undoubtedly, some of these doped
polythiophene-chain materials will superconduct. The main question is what
maximum value of $T_c$ can be attained in these organic compounds.

From the discussions in this section, one can make an important conclusion,
namely that, {\em for the occurrence of superconductivity at room
temperature, the onset of long-range phase coherence will be the bottleneck,
not the quasiparticle pairing}.

In the case of graphene or other two-dimensional materials, an alternative way
to create a room-temperature superconductor can be suggested. One can dope
graphene directly by species which become antiferromagnetically ordered in the
doped state above room temperature, and induce the states at the Fermi level.

\section{Conclusions}

This chapter presents analysis of experimental data. On the basis of this
analysis, it is possible to draw conclusions about components and the structure
of a superconductor with a critical temperature which may exceed the room
temperature. The two essential components of a promising room-temperature
superconductor are large organic molecules (polymers, tissues) and
atoms/molecules which are magnetic in the intercalated state. To reach this
conclusion, one does not require knowledge of the mechanism of
room-temperature superconductivity. However, to synthesize a room-temperature
superconductor is a very difficult task and, in a first approximation, equivalent to
the task of finding a needle in a haystack. Nevertheless, the importance of this
chapter is in that it indicates in which ``haystack'' to search.

\section{A note about the mechanism of superconductivity in cuprates}

It is possible that in cuprates there are two more or less
independent processes, leading to the occurrence of high-$T_c$
superconductivity. One of them is the formation of incoherent
electron (hole) pairs, and the second process is the Bose-Einstein
condensation of magnetic excitations, for example, magnons.
Recently, the Bose-Einstein condensation of magnons was observed in
a number of antiferromagnetic compounds \cite{22}. If the Cooper
pairs are coupled to magnetic excitations, then the pairs can adjust
their phases, i.e. establish the long-range phase coherence, through
the long-range phase coherence of the magnon Bose-Einstein
condensate. The Cooper pairs can be coupled to magnetic excitations
through the amplitude of their wavefunctions, or only through the
phase, or through both.


\begin{thebibliography}{21}

\bibitem{1} Kamerlingh Onnes, H. \emph{Commun. Phys. Lab. Univ. Leiden} 1911,
{\bf 124c}.

\bibitem{2} Mourachkine, A. {\it Room-Temperature Superconductivity};
Cambridge International Science Publishing: Cambridge, 2004, pp 12,
71, 81, 129, 136, 285, 107, 293 (also
http://arxiv.org/ftp/cond-mat/papers/0606/0606187.pdf).

\bibitem{3} Mourachkine, A.  {\it High-Temperature Superconductivity
in Cuprates: The Nonlinear Mechanism and Tunneling Measurements};
Kluwer Academic Publishers: Dordrecht, 2002, pp 247, 246,

\bibitem{4}  Geballe, T. H. \emph{Science} 1993, {\bf 259}, 1550.

\bibitem{5} Bardeen, J.; Cooper, L. N.; Schrieffer, J. R. \emph{Phys.
Rev.} 1957, {\bf 108}, 1175.

\bibitem{6}
Weller, T. E.; Ellerby, M.; Saxena, S. S.; Smith, R. P.; Skipper N.
T. \emph{Nature Physics} 2005, {\bf 1}, 39.

\bibitem{7} Homes, C. C.;  Dordevic, S. V.; Strongin, M.; Bonn, D. A.; Liang Ruixing;
Hardy W. N.;  Komiya Seiki; Ando Y.; Yu, G.; Kaneko, N.; Zhao, X.;
Creven, M.; Basov, D. N.; Timusk, T. \emph{Nature} 2004, {\bf 430},
539.

\bibitem{8} Little, W. A. \emph{Phys. Rev.} 1964, {\bf 134}, A1416.

\bibitem{9}
Kresin, V. Z.; Litovchenko, V. A.; Panasenko, A. G. \emph{J. Chem.
Phys.} 1975, {\bf 63}, 3613.

\bibitem{10}
Andriotis, A. N.; Menon, M.; Sheetz, R. M.; Chernozatonskii, L.
\emph{Phys. Rev. Lett.} 2003, {\bf 90}, 026801.

\bibitem{11}
Zhao, G.-m. preprint 2000, cond-mat/0208200.

\bibitem{12}
Heeger, A. J.; Kivelson, S.; Schrieffer, J. R.; Su, W.-P. \emph{Rev.
Mod. Phys.} 1988, {\bf 60}, 781.

\bibitem{13}
Friend R. H.; Gymer R. W.; Holmes A. B.; Burroughes J. H.; Marks R.
N.; Taliani C.; Bradley D. D. C.; Dos Santos D. A.; Br\'edas J. L.;
L\"ogdlund M.; Salaneck W. R. \emph{Nature} 1999, {\bf 397}, 121.

\bibitem{14}
Davydov, A. S. {\it Solitons in Molecular Systems}; \emph{Naukova
Dumka}: Kiev, 1988 (in Russian), p. 103.

\bibitem{15} Davydov, A. S. \emph{Phys. Rep.} 1990, {\bf 190}, 191.

\bibitem{16}
Davydov, A. S. {\it Solitons in Molecular Systems};
Kluwer Academic Publishers: Dordrecht, 1991, p. 124.

\bibitem{17}
Pullman, B.; Pullman, A. {\it Quantum Biochemistry}; Interscience: New York,
1963.

\bibitem{18}
Clardy, J.; Walsh, C. \emph{Nature} 2004, {\bf 432}, 829.

\bibitem{19}
{\it Graphite Intercalation Compounds}, Tanuma, S. and
Kamimura, H.; Ed.; World Scientific: Singapore, 1985.

\bibitem{20}
Kopelevih, Y.; Esquinazi, P.; Torres, J. H.; da Silva, R. R.; Kempa, H.; Mrowka, F.;
Ocana, R. preprint 2002 (a chapter in a book), cond-mat/0209442,
and references therein.

\bibitem{21}
Novoselov, K. S.; Jiang, D.; Booth, T.; Khotkevich, V. V.; Morozov,
S. M.; Geim, A. K. \emph{PNAS} 2005, {\bf 102}, 10451.

\bibitem{22} See, for example, Ruegg, Ch.; Cavadini, N.; Furrer, A.; Gudel, H.-U.;
Kramer, K.; Mutka, H.; Wildes, A.; Habicht, K.; Vorderwisch, V.
\emph{Nature} 2003, {\bf 423}, 62.


\end{thebibliography}
\end{document}